\documentclass[prl,showpaces,preprintnumbers,amsmath,amssymb,twocolumn,superscriptaddress]{revtex4}
\usepackage{graphicx}
\usepackage{dcolumn}
\usepackage{bm}
\usepackage{amssymb}
\usepackage{amsmath}
\usepackage{epstopdf}
\usepackage[dvipdfm, pdfstartview=FitH, CJKbookmarks=true, bookmarksnumbered=true, bookmarksopen=true, colorlinks, pdfborder=001, linkcolor=blue, anchorcolor=blue, citecolor=blue]{hyperref}

\begin{document}


\title{Weak Interband-Coupling Superconductivity in the Filled Skutterudite LaPt$_{4}$Ge$_{12}$}

\author{J. L. Zhang} \affiliation{Center for Correlated Matter and Department of Physics, Zhejiang University, Hangzhou, Zhejiang 310058, China}
\affiliation{High Magnetic Field Laboratory, Chinese Academy of Sciences, Hefei 230031, China}
\author{G. M. Pang} \affiliation{Center for Correlated Matter and Department of Physics, Zhejiang University, Hangzhou, Zhejiang 310058, China}
\author{L. Jiao} \affiliation{Center for Correlated Matter and Department of Physics, Zhejiang University, Hangzhou, Zhejiang 310058, China}
\author{M. Nicklas} \email{nicklas@cpfs.mpg.de}\affiliation{Max Planck Institute for Chemical Physics of Solids, N\"{o}thnitzer Stra{\ss}e 40, 01187 Dresden, Germany}
\author{Y. Chen} \affiliation{Center for Correlated Matter and Department of Physics, Zhejiang University, Hangzhou, Zhejiang 310058, China}
\author{Z. F. Weng} \affiliation{Center for Correlated Matter and Department of Physics, Zhejiang University, Hangzhou, Zhejiang 310058, China}
\author{W. Schnelle} \affiliation{Max Planck Institute for Chemical Physics of Solids, N\"{o}thnitzer Stra{\ss}e 40, 01187 Dresden, Germany}
\author{A. Leithe-Jasper} \affiliation{Max Planck Institute for Chemical Physics of Solids, N\"{o}thnitzer Stra{\ss}e 40, 01187 Dresden, Germany}
\author{A.~Maisuradze} \affiliation{Laboratory for Muon Spin Spectroscopy, Paul Scherrer Institut, CH-5232 Villigen PSI, Switzerland} \affiliation{Physik-Institut der Universit\"{a}t Z\"{u}rich, Winterthurerstrasse 190, CH-8057 Z\"{u}rich, Switzerland}
\author{C.~Baines} \affiliation{Laboratory for Muon Spin Spectroscopy, Paul Scherrer Institut, CH-5232 Villigen PSI, Switzerland}
\author{R.~Khasanov} \affiliation{Laboratory for Muon Spin Spectroscopy, Paul Scherrer Institut, CH-5232 Villigen PSI, Switzerland}
\author{A.~Amato} \affiliation{Laboratory for Muon Spin Spectroscopy, Paul Scherrer Institut, CH-5232 Villigen PSI, Switzerland}
\author{F. Steglich } \affiliation{Center for Correlated Matter and Department of Physics, Zhejiang University, Hangzhou, Zhejiang 310058, China}
\affiliation{Max Planck Institute for Chemical Physics of Solids, N\"{o}thnitzer Stra{\ss}e 40, 01187 Dresden, Germany}
\author{R. Gumeniuk} \affiliation{Max Planck Institute for Chemical Physics of Solids, N\"{o}thnitzer Stra{\ss}e 40, 01187 Dresden, Germany}\affiliation{Institut f{\"u}r Experimentelle Physik, TU Bergakademie Freiberg, Leipziger Stra{\ss}e 23, 09596 Freiberg, Germany}
\author{H. Q. Yuan} \email{hqyuan@zju.edu.cn} \affiliation{Center for Correlated Matter and Department of Physics, Zhejiang University, Hangzhou, Zhejiang 310058, China}
\affiliation{Collaborative Innovation Center of Advanced Microstructures, Nanjing 210093, China}
\date{\today}

\begin{abstract}

The superconducting pairing state of LaPt$_{4}$Ge$_{12}$ is studied by measuring the magnetic penetration depth $\lambda(T,B)$ and the superfluid density $\rho_s(T)$ using a tunnel-diode-oscillator (TDO)-based method and by transverse field muon-spin rotation ($\mu$SR) spectroscopy. $\lambda(T)$ follows an exponential-type temperature dependence at $T\ll T_{c}$, but its zero-temperature value $\lambda(0)$ increases linearly with magnetic field. Detailed analyses demonstrate that both $\lambda(T)$ and the corresponding $\rho_{s}(T)$, measured in the Meissner state by the TDO method are well described by a two-gap $\gamma$ model with gap sizes of $\Delta_1(0)=1.31k_{B}T_c$ and $\Delta_2(0)=1.80k_{B}T_c$ and a very weak interband coupling. In contrast, $\rho_s(T)$, derived from the $\mu \rm{SR}$ data taken in a small field, can be fitted by a single-gap BCS model with a gap close to $\Delta_2(0)$. We conclude that LaPt$_{4}$Ge$_{12}$ is a marginal two-gap superconductor and the small gap $\Delta_1$ seems to be destroyed by a small magnetic field. In comparison, in PrPt$_4$Ge$_{12}$ the $4f$-electrons may enhance the interband coupling and, therefore, give rise to more robust multiband superconductivity.

\end{abstract}


\maketitle

The discovery of superconductivity (SC) in a series of Pr-based skutterudite compounds, i.e., Pr\textit{T}$_{4}$\textit{X}$_{12}$ (\textit{T}=Fe, Ru, Os, and \textit{X}=pnicogen), has attracted considerable interests. Despite of intensive investigations in the physical properties of these compounds, their superconducting order parameter and, therefore, the pairing mechanism remain highly controversial \cite{Sales}: nodal SC was evidenced in some of these compounds, while conventional BCS SC was recognized in others. Even for the same compound, most prominently, the heavy-fermion superconductor PrOs$_{4}$Sb$_{12}$, the gap symmetry is still under debate. Here, the early measurements provided evidence of point nodes with a possible triplet pairing state for PrOs$_{4}$Sb$_{12}$ \cite{Higemoto07, Izawa, Chia}. With improving sample homogeneity and lowering the accessible temperatures, recent measurements of thermal conductivity demonstrated that PrOs$_{4}$Sb$_{12}$ belongs to multiband superconductors \cite{Seyfarth06, Hill}. Exploration of other filled skutterudite superconductors may help to elucidate the superconducting pairing state and the question of its universality.

A family of Ge-based skutterudites with $T_{c}$ ranging from 5 to 8.3 K , i.e., \textit{M}Pt$_4$Ge$_{12}$ (\textit{M}=Sr, Ba, La, Pr), was discovered a few years ago \cite{Gumeniuk, Bauer2007, Kaczorowski08, Bauer08Th}. Resembling those of PrOs$_{4}$Sb$_{12}$, zero-field $\mu$SR experiments showed evidence of time-reversal symmetry (TRS) breaking in the superconducting state of PrPt$_{4}$Ge$_{12}$, but not for LaPt$_{4}$Ge$_{12}$ \cite{Maisuradze09PRL, Maisuradze10PRB,Zhang15}. However, more recent experiments suggested multi-gap SC for PrPt$_{4}$Ge$_{12}$ \cite{Nakamura12, Chandra12, Zhang13}. Furthermore, a coherence peak was revealed in the nuclear spin-lattice relaxation rate $1/T_1$ just below $T_c$ for PrPt$_{4}$Ge$_{12}$, which is unexpectedly suppressed in LaPt$_{4}$Ge$_{12}$ \cite{Kanetake09}. However, the weak variation of $T_c$ and of the specific heat jump at $T_c$ with the Pr-content $x$ in La$_{1-x}$Pr$_x$Pt$_4$Ge$_{12}$ suggests that the order parameters of the two end members are compatible \cite{Maisuradze10PRB}. Thus, LaPt$_{4}$Ge$_{12}$ is not a simple reference compound of PrPt$_{4}$Ge$_{12}$, as seen in many heavy fermion systems. In order to clarify the superconducting order parameters in LaPt$_{4}$Ge$_{12}$, further investigations are badly needed. Furthermore, a comparative study of PrPt$_{4}$Ge$_{12}$ and LaPt$_4$Ge$_{12}$ could shed new light on the role of Pr-4$f$ electrons in the formation of SC and, therefore, provide a more general picture of pairing states in skutterudite superconductors.

In this Letter, we report a comprehensive study of the magnetic penetration depth $\lambda(T, B)$ and superfluid density $\rho_s(T)$ of LaPt$_{4}$Ge$_{12}$ by using a $\mu$SR experiment and a TDO-based technique. The London penetration depth $\lambda_L(T)$, measured by the TDO method, follows exponential-type behavior in the low-temperature limit, suggesting nodeless superconductivity for LaPt$_{4}$Ge$_{12}$. The zero-temperature penetration depth $\lambda_m(B)$, determined from the $\mu$SR experiments, follows linear field dependence, giving evidence of multigap superconductivity. The superfluid density $\rho_{s}^{\textrm{TDO}}(T)$, converted from $\lambda_L(T)$, is best described by a two-gap $\gamma$-model, with gap sizes of $\Delta_1(0)=1.31 k_{B}T_{c}$ and $\Delta_2(0) =1.80 k_{B}T_{c}$ at zero temperature. However, the superfluid density $\rho_s^{\mu\textrm{SR}}(T)$ derived from the $\mu$SR experiment with a transverse field of 75~mT is fitted by a conventional BCS model with a single gap $\Delta(0)\approx\Delta_2(0)$. We argue that marginal two-gap SC exists in LaPt$_{4}$Ge$_{12}$ and also some other skutterudite superconductors. Due to the very weak interband coupling in LaPt$_{4}$Ge$_{12}$, the gap with smaller $\Delta/k_{B}T_{c}$ is easily destroyed by weak magnetic fields (as used in our $\mu$SR and also in some experimental probes) or other perturbations.

\begin{figure}[tb]
\includegraphics[width=8cm]{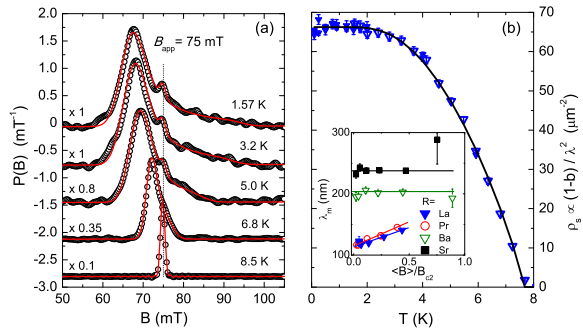}
\caption{(Color online) (a) Fourier transforms of the $\mu$SR time spectra $P(B)$ of LaPt$_4$Ge$_{12}$ at various temperatures, with an applied field of $B_\mathrm{app} = 75$~mT. For better visualization the data at each temperature were scaled and shifted vertically. The solid lines are fits to the data. (b) Temperature dependence of the superfluid density
$\rho_s^{\mu \textrm{SR}} \propto (1-b)/\lambda_m^{2}$. The solid line shows a BCS fit. Inset: magnetic penetration depth $\lambda_{m}$ as a function of normalized field $\langle B\rangle/B_{\rm c2}$ in \textit{M}Pt$_{4}$Ge$_{12}$ $M={\rm La}$, Ba \cite{Maizuradze2012}, Sr, and Pr \cite{Maisuradze09PRL}.}
 \label{fig:Spectra}
\end{figure}

The $\mu$SR experiments were performed at the Paul Scherrer Institute (Switzerland) at the $\pi$M3 beam line on the GPS spectrometer down to 1.5 K and at the LTF spectrometer down to $T\approx0.02$~K on polycrystalline LaPt$_4$Ge$_{12}$ samples \cite{Gumeniuk}. The samples were cooled in $B_{\rm app}=75$~mT from above $T_c$ down to 1.5~K, and then measured as a function of temperature. At $T=1.5$~K we also collected data in magnetic fields. Typical counting statistics were $7-8\times10^6$ positron events per each data point. High-quality single crystals \cite{Gumeniuk2010ZK}, were utilized for the London penetration depth study by a TDO-based method. The $ac$ field induced by the coil of the experimental setup was less than 2~$\mu$T, which is much smaller than the lower critical field of LaPt$_4$Ge$_{12}$, guaranteeing that the sample was in the Meissner state. Typical dimensions of the plate-like samples were about $(0.6-1.0)\times(0.6-1.0)\times0.2 \rm{~mm}^{3}$. The change of the London penetration depth $\Delta\lambda_L(T)$ is proportional to the frequency shift, i.e.\ $\Delta\lambda_L(T)=G\cdot\Delta f(T)$, where the $G$ factor is solely determined by the sample and coil geometries \cite{Prozorov 11}.

\begin{figure}[b]\centering
 \includegraphics[width=8cm]{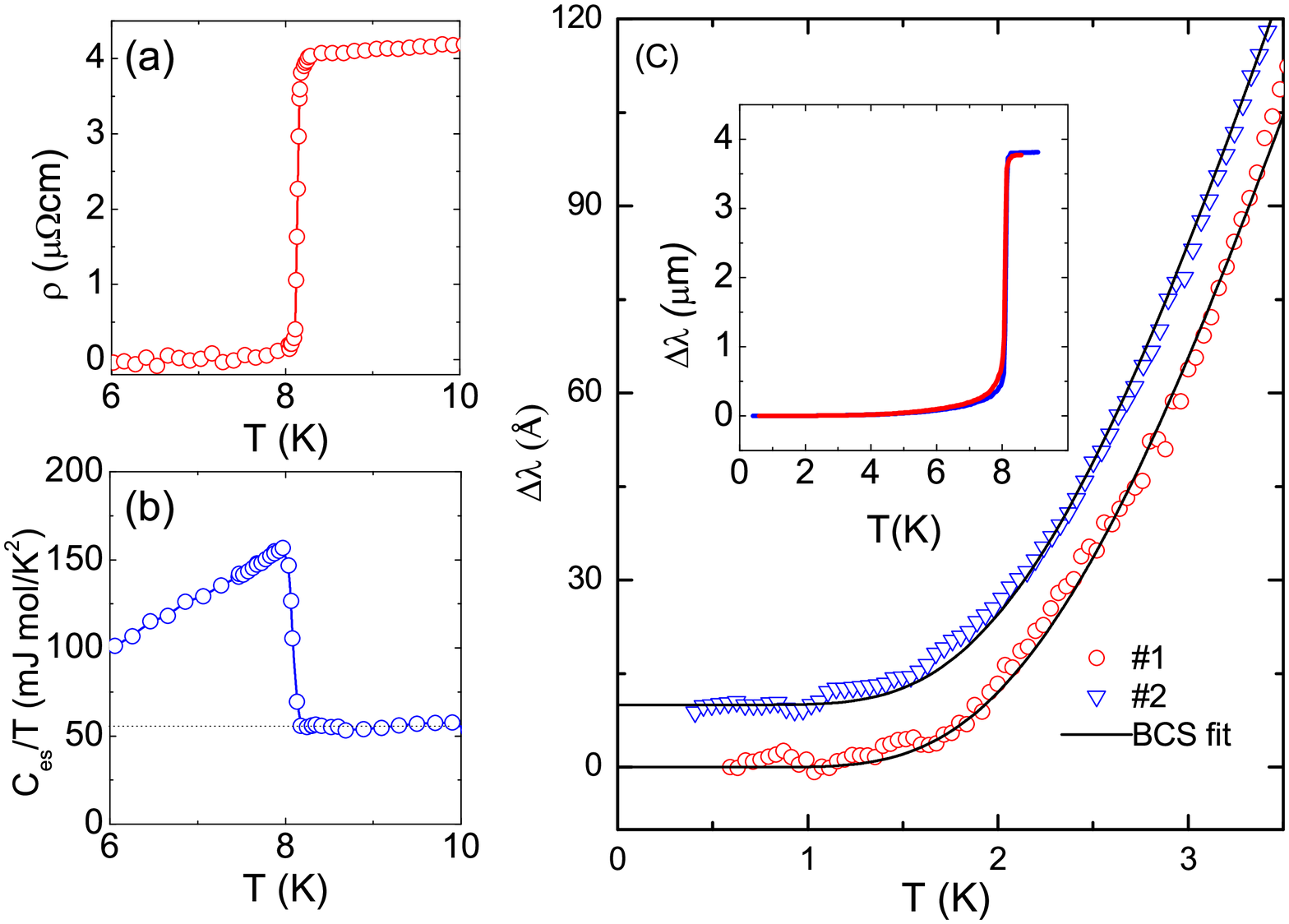}
\caption{Temperature dependence of (a) the electrical resistivity $\rho(T)$, (b) the electronic specific heat, and (c) the London penetration depth $\Delta\lambda_L(T)$ for LaPt$_{4}$Ge$_{12}$ single crystals. The inset shows $\Delta\lambda_L(T)$ over a wide temperature
range. For clarity, the $\Delta\lambda_L(T)$ data are shifted by 10
${\rm{\AA}}$ for the two samples ($\#$1 and $\#$2). The solid lines present a fit of the BCS model at $T\ll T_C$ to the experimental $\Delta\lambda_L(T)$.}\label{fig.2}
\end{figure}

Figure~\ref{fig:Spectra}(a) shows the Fourier transforms of the $\mu$SR time spectra $P(B)$ of LaPt$_4$Ge$_{12}$ at various temperatures for $B_{\rm app}=75$~mT. Above $T_{c} = 8.3$~K, a narrow and sharp peak is visible in $P(B)$ as it is expected for a weak nuclear depolarization. With lowering the temperature, the peak is strongly broadened and becomes asymmetric, indicating a field distribution within
the well arranged flux-line lattice (FLL). These features point to isotropic SC and weak pinning of the FLL in the polycrystalline samples. In this context, we adopt the exact solution of the Ginzburg-Landau equations with the method suggested by Brandt to analyze the data \cite{Brandt97_NGLmethod,Brandt03}. The
complete analysis of the data follows the procedure as described in Ref.~\onlinecite{Maizuradze2012}. Accordingly, we used the spatial magnetic field distribution $B({\bf r}) = B({\bf r}, \lambda_m, \xi, \langle B \rangle)$ within the unit cell of the FFL, where $\lambda_m$ is the magnetic penetration depth and $\xi$ is the coherence length. The data are well fitted by using the theoretical polarization function $P(t)$ given in Ref.~\onlinecite{Maizuradze2012}, as denoted by the solid lines in Fig.~\ref{fig:Spectra}(a).

The mean value of the superfluid density can be expressed by $\rho_s^{\mu \textrm{SR}}\propto(1-b)/\lambda_m^2$, where $b=\langle B \rangle/B_{c2}(0)$ is the reduced field. In Fig.~\ref{fig:Spectra}(b), we plot $\rho_s^{\mu \textrm{SR}}$ derived from the $\mu$SR data as a function of temperature, which can be nicely fitted by the conventional $s$-wave BCS model with a gap amplitude of $\Delta(0)=1.9 k_{B}T_{c}$. The detailed analysis of the superfluid density will be presented in comparison with the data obtained from the TDO measurements (see below).

The field dependence of the magnetic penetration depth provides an alternative way to look into the low-energy excitations. For nodal or multi-gap superconductors, the penetration depth $\lambda_m$ depends on the applied magnetic field \cite{Amin00}, but it hardly varies with  field for conventional s-wave superconductors \cite{LandauKeller07}. In the inset of Fig.~\ref{fig:Spectra}b we show the field dependence of $\lambda_m$ obtained at 1.5 K for several superconducting skutterudites based on the Pt$_4$Ge$_{12}$ framework. In order to avoid correlation effects between $\lambda$ and $\xi$, we fixed the values of the coherence length $\xi=15.0$, 19.7, 14.2, and 27.0 nm for the La-, Sr-, Pr-, and BaPt$_4$Ge$_{12}$ skutterudites, respectively \cite{Maisuradze08}. These values are obtained from the corresponding upper critical fields $B_{\rm c2}$ at 1.5 K using the Ginzburg-Landau relation $B_{\rm c2} = \Phi_0/2\pi\xi^2$. $\lambda_m$ of SrPt$_4$Ge$_{12}$ and BaPt$_4$Ge$_{12}$ are independent of the applied field, as expected for s-wave SC with an isotropic gap (see Fig.~\ref{fig:Spectra}b). In contrast, both LaPt$_4$Ge$_{12}$ and PrPt$_4$Ge$_{12}$ display a linear dependence of $\lambda_m$ on the magnetic field with a similar slope, which
remarkably resembles that of MgB$_{2}$ \cite{Ohishi}, a prototype of two-band superconductors. Such behavior is compatible with the multi-band SC in PrPt$_4$Ge$_{12}$ \cite{Chandra12,Nakamura12,Zhang13}, but contradicts to the above-mentioned single-band BCS SC for LaPt$_4$Ge$_{12}$.

\begin{figure}[b]\centering
 \includegraphics[width=8cm]{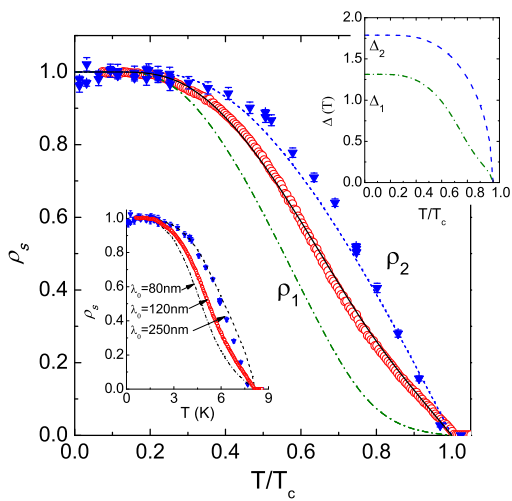}
\caption{The normalized superfluid density $\rho_s(T)$ for LaPt$_4$Ge$_{12}$. The circles and triangles represent the experimental data obtained by using the TDO method ($\rho_{s}^{\rm TDO}$) and the TF-$\mu$SR measurements ($\rho_{s}^{\mu \rm{SR}}$), respectively. The solid line is a fit based on the $\gamma$ model; the derived partial superfluid density $\rho_{1}(T)$ and $\rho_{2}(T)$ are shown by the dashed and dash-dotted lines, respectively. Lower
inset: The superfluid density $\rho_s^{\textrm{TDO}}(T)$ calculated with $\lambda_{0}=80$ nm, 120 nm and 250 nm. Upper inset: Temperature dependence of the gap amplitudes, $\Delta_{1}(T)$ and $\Delta_{2}(T)$.}\label{fig.3}
\end{figure}

\begin{figure}[b]\centering
 \includegraphics[width=8cm]{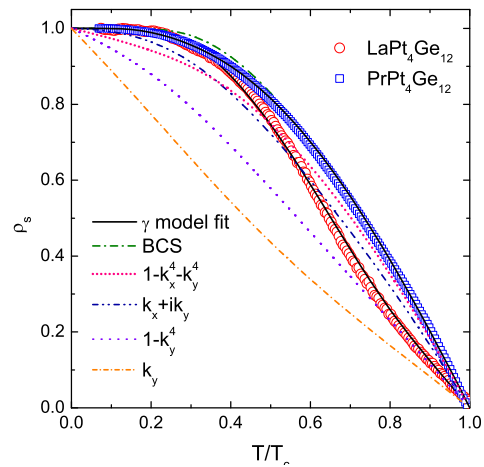}
\caption{The normalized superfluid density, $\rho_s^{\textrm{TDO}}(T)$, for LaPt$_{4}$Ge$_{12}$ and PrPt$_{4}$Ge$_{12}$. The symbols and solid lines represent the experimental data and their corresponding fits using the two-band $\gamma$-model. For comparison, we also fit the $\rho_s^{\rm TDO}(T)$ of LaPt$_4$Ge$_{12}$ using various gap functions allowed by the crystal symmetry \cite{Zhang13}; the gap functions are denoted in the figure.}\label{fig.4}
\end{figure}

In order to elucidate the above contradictory behaviors, we measured the London penetration depth using the TDO-based method for single crystalline LaPt$_4$Ge$_{12}$. The sharp superconducting transitions at $T_{c}=8.2$~K, observed in both the electrical resistivity $\rho(T)$ and the electronic specific heat $C_{es}(T)/T$ as shown in Fig.~\ref{fig.2}a and \ref{fig.2}b, together with the low residual resistivity, confirm the high quality of the LaPt$_4$Ge$_{12}$ crystals. Figure~\ref{fig.2}c presents the London penetration depth, $\Delta\lambda_L(T)$, for two samples from the same batch. Obviously, $\Delta\lambda_L(T)$ is well reproducible among different samples, showing nearly temperature-independent behavior below 1.5~K. This provides further evidence of a nodeless SC gap in LaPt$_4$Ge$_{12}$. According to the BCS model, the penetration depth $T\ll T_c$ can be expressed by \cite{BCS Prozorov}: $\Delta\lambda(T)\approx\lambda(0)\sqrt{\frac{\pi\Delta(0)}{2T}}\exp({-\frac{\Delta(0)}{T}})$, where $\lambda(0)$ and $\Delta(0)$ represent the penetration depth and gap amplitude at zero temperature, respectively. Here we take the value of $\lambda(0)\approx 120$~nm from the $\mu$SR experiments. Usually, the parameter $\Delta(0)$ varies with changing the temperature range for the fit, but becomes saturated at sufficiently low temperatures. The solid lines in Fig.~\ref{fig.2}c represent the best fits to the BCS model in the low-temperature limit, giving $\Delta(0)=1.39 k_{B}T_c$ for sample $\#$1 and $\Delta(0)=1.34 k_{B}T_c$ for sample $\#$2. The so-derived $\Delta(0)$ is significantly smaller than the BCS value of $\Delta(0)=1.76 k_{B}T_c$, which is usually attributed to the opening of a secondary small gap at low temperature or the anisotropic effects.

In Fig.~\ref{fig.3}, we plot the normalized superfluid density $\rho_s^{\textrm{TDO}}(T) =\lambda^2(0)/\lambda_L^2(T)$ for LaPt$_4$Ge$_{12}$, which shows a concave curvature near $T_c$. Such a behavior might result from the underestimation of $\lambda(0)$. As a comparison, the lower inset of Fig.~\ref{fig.3} shows $\rho_{s}^{\textrm{TDO}}(T)$ calculated by using different values of $\lambda(0)$. Indeed, the concave curvature of $\rho_{s}^{\textrm{TDO}}(T)$ weakens with increasing $\lambda(0)$; $\rho_{s}^{\textrm{TDO}}(T)$ resembles the $\mu \rm{SR}$ data for $\lambda(0)=250$~nm, but still shows clear deviations near $T_c$.  It is noted that our samples have a rather regular geometry and the total uncertainties from the $\lambda(0)$ value and the G$-$factor in the TDO measurements should be limited to about 20\%. Thus, such a large enhancement of $\lambda(0)$ is also far beyond the uncertainties from our experiments. On the other hand, a similar concave behavior was previously observed in some multiband superconductors with weak interband coupling \cite{Kogan}. Considering the fact that LaPt$_{4}$Ge$_{12}$ possesses multiple sheets of Fermi surface \cite{Gumeniuk}, it is reasonable to analyze the $\rho_{s}^{\textrm{TDO}}(T)$ in terms of multiband SC.

According to the $\gamma$ model \cite{Kogan}, which applies generally to clean two-band superconductors, the total superfluid density can be written as:
$\rho_{s}(T)=\gamma\rho_{1}(\Delta_{1}(0),T)+(1-\gamma)\rho_{2}(\Delta_{2}(0),T)$,
where $\gamma$ is the relative weight of contributions from the gap $\Delta_{1}$. For a two-band system with known density of states $n_{\mu}$ and Fermi velocities $\upsilon_{\mu}$ ($\mu=1,2$), $\gamma$ is determined by $\gamma=n_1\langle \upsilon_1^2 \rangle/(n_1\langle \upsilon_1^2 \rangle+n_2\langle \upsilon_2^2 \rangle$). Here $\langle \upsilon_{\mu}^2\rangle$ represents the average of the squared Fermi velocities over the corresponding band. In LaPt$_4$Ge$_{12}$, several energy bands cross the Fermi energy \cite{Gumeniuk}. As an approximation, we group the major energy bands into two effective bands according to the values of Fermi velocity in each band, which gives $n_{1}=0.12$, $n_2=0.88$, and $\upsilon_1/\upsilon_2\sim2.3$. Thus, a value of $\gamma=0.42$ is estimated. The partial superfluid density $\rho_{\mu}$ and the energy gap $\Delta_{\mu}$ are calculated self-consistently within the quasiclassic Eilenberger equations. Following the method described in Ref. \onlinecite{Kogan}, we fit the experimental $\rho_s^{\textrm{TDO}}(T)$ with adjusted parameters of $\gamma$ and $\lambda_{\mu\nu}$; here $\lambda_{\mu\nu}$ are the interband and intraband pairing couplings. The best-fit results, with parameters $\lambda_{11}$=0.63, $\lambda_{12}=0.0025$, and $\lambda_{22}=0.20$, are shown in Fig.~\ref{fig.3}, which overlaps nicely with the experimental results. The derived values of $n_{1}=0.23$ and $\gamma=0.4$ are compatible with those estimated directly from the band structure calculations (see above). Such a fit gives $\Delta_1(0)=1.31 k_{B}T_{c}$ and $\Delta_2(0)=1.80 k_{B}T_{c}$; the small gap $\Delta_1 (0)$ agrees well with that obtained from the exponential fits of $\Delta\lambda_L(T)$ at $T\ll T_c$, where $\Delta(0)=1.39 k_{B}T_c$. As shown in the upper inset of Fig.~\ref{fig.3}, $\Delta_1(T)$ exhibits non-BCS behavior; the rapid vanishing of $\Delta_1(T)$ leads to the concave curvature in $\rho_s^{\textrm{TDO}}(T)$ near $T_c$.

Remarkably, the so-derived partial superfluid density $\rho_2(T)$ is nearly identical to $\rho_s^{\mu \textrm{SR}}(T)$ (see Fig.~\ref{fig.3}). Such an agreement suggests that the small energy gap is readily destroyed by the small magnetic field applied in the $\mu$SR experiments. In a two-band superconductor with a vanishing interband coupling $\lambda_{12}$, it was theoretically shown that the small gap may be suppressed at a largely reduced effective upper-critical field \cite{Tewordt}. The so-called ``virtual" upper-critical field $B_{c2}^s$, above which the vortex cores overlap and thus drive the majority of the electrons in the associated band normal, can be estimated by $B_{c2}^{s}(0)\sim B_{c2}(0)[\Delta_s(0) \upsilon_{Fl}/\Delta_l(0) \upsilon_{Fs}]^2$. For LaPt$_4$Ge$_{12}$, a value of $B_{c2}^s(0)\sim 150$~mT is estimated by using the parameters derived above. It should be noted that the so-derived $B^{s}_{c2}(0)\sim$150 mT is likely overestimated, because the above formula was derived for MgB$_{2}$ which has a larger interband coupling \cite{Tewordt}. The small $B_{c2}^s(0)$ may well describe the discrepancy of the superfluid density $\rho_s(T)$ determined from the TDO and the $\mu$SR experiments. In the TDO measurements, the sample is in the Meissner state, which allows us to detect both the large and the small gaps. On the other hand, the TF-$\mu$SR measurements were carried out in a field of 75~mT. In this case, the small gap $\Delta_1$ is already significantly suppressed and $\rho_{s}^{\mu \textrm{SR}}(T)$ is governed by the large gap, leading to a single-band BCS behavior with a large energy gap $\Delta_2$.  Our analysis suggests that LaPt$_4$Ge$_{12}$ is a marginal two-band superconductor; the signatures of the small gap being easily extinguished by magnetic fields or impurity scattering due to the weak interband coupling.

It was previously shown that the sister compound, PrPt$_{4}$Ge$_{12}$, is also a two-gap superconductor featuring a relatively small admixture of a second gap with a small $\Delta_0/k_{B}T$ \cite{Zhang13}. As a comparison, we plotted the superfluid densities $\rho_s^{\textrm{TDO}}(T)$ of LaPt$_{4}$Ge$_{12}$ and PrPt$_{4}$Ge$_{12}$ in Fig.~\ref{fig.4}, which show different curvatures near $T_c$. In Fig.~\ref{fig.4}, we reanalyzed the $\rho_s^{\textrm{TDO}}$ of PrPt$_{4}$Ge$_{12}$ in terms of the $\gamma$-model. The best fit gives intraband coupling parameters of $\lambda_{11}$=0.59 and $\lambda_{22}$=0.18, which are consistent with those of LaPt$_{4}$Ge$_{12}$. On the other hand, the derived interband coupling $\lambda_{12}$=0.07 is more than one order of magnitude larger than that of LaPt$_4$Ge$_4$. The enhanced interband coupling leads to a more robust small energy gap, showing the same type of temperature dependence for the two energy gaps \cite{Zhang13}. For PrPt$_{4}$Ge$_{12}$, the gap parameters derived from the $\gamma$-model [$\Delta_1(0)=0.88 k_{B}T_{c}$, $\Delta_2(0)=2.17 k_{B}T_{c}$ and $\gamma=0.15$] are compatible with those fitted by the $\alpha$-model \cite{Zhang13}. These results suggest that the Pr-4$f$ electrons might not play a crucial role in the formation of SC, but enhance the interband coupling in the \textit{M}Pt$_{4}$Ge$_{12}$ compounds. This finding is consistent with the local-density approximation calculations, which show similar electronic structures for the two compounds, with the density of states at the Fermi level mainly contributed by the Pt and Ge states \cite{Gumeniuk}. Following the same methods used in Ref.~\onlinecite{Zhang13} for PrPt$_4$Ge$_{12}$, we also fit $\rho_s^{\textrm{TDO}}(T)$ of LaPt$_4$Ge$_{12}$ with various gap functions (see Fig.~\ref{fig.4}). One recognizes that all the other models give only a poor fit to the experimental data, further supporting the two-gap SC of LaPt$_4$Ge$_{12}$.

Our results may have broader implications on reconciling the diverse behaviors of several multiband superconductors, in particular those with a weak interband coupling. A number of filled skutterudite SCs have been considered as two-band SCs on the basis of various experiments \cite{Seyfarth06, Hill, Adroja}, while in others the data are not convincing \cite{Shu09, Chia2004}. Other similar examples also include V$_3$Si, La$_2$C$_3$, MgCNi$_3$ and other materials, where different experimental probes are not yet conclusive on the order parameter symmetry \cite{Zehetmayer}. Focusing on our results, we find that, in LaPt$_{4}$Ge$_{12}$, the smaller gap is subtle and could be easily suppressed by magnetic fields due to the weak interband couplings. Seyfarth \textit{et al.}\ [\onlinecite{Seyfarth06}] pointed out that crystal inhomogeneities may also prevent from observing the small energy gap.

In summary, we have investigated the superconducting order parameter of LaPt$_{4}$Ge$_{12}$ by measuring the penetration depth and the superfluid density with two distinct methods. We find that the London penetration depth $\triangle\lambda_L(T)$ and the superfluid density $\rho_s(T)$, derived from the Meissner state using the TDO measurements, can be consistently described by the two-band $\gamma$-model with $\Delta_1(0)=1.31 k_{B}T_{c}$ and $\Delta_2(0)=1.80 k_{B}T_{c}$. The small gap is likely destroyed by a small magnetic field as it is applied in the $\mu$SR experiments. Our results suggest that LaPt$_2$Ge$_{12}$ is a two-band superconductor where the critical field for suppressing the small gap is largely reduced due to the extraordinarily weak interband coupling. The presence of 4$f$-electrons in the Pr-based skutterudite superconductors may enhance the interband coupling and, therefore, give rise to more robust multiband behavior.

\begin{acknowledgments}

We would like to acknowledge useful discussions with H.\ Pfau, H.\ Rosner, Yu.\ Grin, C.\ Cao and X.\ Lu. This work was partially supported by the National Basic Research Program of China (No.\ 2011CBA00103), the National Natural Science Foundation of China (No.\ 11474251 and No.\ 11174245) and the Fundamental Research Funds for the Central Universities, the Max Planck Society under the auspices of the Max Planck Partner Group of the Max Planck Institute for Chemical Physics of Solids, Dresden. The work performed at the Swiss Muon Source (S$\mu$S), Paul Scherrer Institut (PSI, Switzerland) was supported by the NCCR program {\it Materials with Novel Electronic Properties} (MaNEP) sponsored by the Swiss National Science Foundation.
\end{acknowledgments}

\end{document}